\documentclass{article}
\usepackage{graphicx} 
\usepackage[margin=1in]{geometry}
\usepackage{cite}
\usepackage{caption}
\usepackage{subcaption}
\usepackage{amsmath}
\usepackage{amssymb}
\usepackage{amsthm}
\usepackage{soul}
\usepackage{placeins}
\usepackage{comment}
\usepackage{xcolor}
\usepackage{authblk}
\usepackage{float}
\definecolor{royalazure}{rgb}{0.0, 0.22, 0.86}
\usepackage[colorlinks=true,linkcolor=blue,citecolor=royalazure
]{hyperref}

    \makeatletter
\def\@fnsymbol#1{\ensuremath{\ifcase#1\or \dagger\or \ddagger\or
   \mathsection\or \mathparagraph\or \|\or **\or \dagger\dagger
   \or \ddagger\ddagger \else\@ctrerr\fi}}
    \makeatother

\begin{document}

\title{Towards quantum-enabled cell-centric therapeutics}

\author[1]{Saugata Basu}
\author[2]{Jannis Born}
\author[3]{Aritra Bose}
\author[4,5]{Sara Capponi}
\author[6]{Dimitra Chalkia}
\author[7,8]{Timothy A Chan}
\author[9]{Hakan Doga}
\author[10]{Frederik F. Fl\"{o}ther}
\author[11,12,13,14]{Gad Getz}
\author[15]{Mark Goldsmith}
\author[9]{Tanvi Gujarati}
\author[3]{Aldo Guzm\'an-S\'aenz}
\author[6]{Dimitrios Iliopoulos}
\author[9]{Gavin O. Jones}
\author[15]{Stefan Knecht}
\author[16]{Dhiraj Madan}
\author[15]{Sabrina Maniscalco}
\author[17]{Nicola Mariella}
\author[3]{Joseph A. Morrone}
\author[18]{Khadijeh Najafi}
\author[2]{Pushpak Pati}
\author[3]{Daniel Platt}
\author[2]{Maria Anna Rapsomaniki}
\author[16]{Anupama Ray}
\author[3]{Kahn Rhrissorrakrai}
\author[18]{Omar Shehab}
\author[19]{Ivano Tavernelli}
\author[9]{Meltem Tolunay}
\author[3]{Filippo Utro}
\author[19]{Stefan Woerner}
\author[17]{Sergiy Zhuk}

\author[9]{Jeannette M. Garcia \thanks{Corresponding author. \{jmgarcia, parida\}@us.ibm.com}\thanks{Authors listed in alphabetical order with exception of corresponding authors.}}
\author[3]{Laxmi Parida$^\dagger$$^\ddag$}

\affil[1]{\small Purdue University, Department of Mathematics, West Lafayette, IN, USA }
\affil[2]{IBM Research, IBM Research Europe, Zurich, Switzerland}
\affil[3]{IBM Research, IBM Thomas J Watson Research Center, Yorktown Heights, NY, USA}
\affil[4]{IBM Research, Almaden Research Center, San Jose, CA, USA}
\affil[5]{Center for Cellular Construction, San Francisco, CA, USA}
\affil[6]{Athos Therapeutics Inc., Los Angeles, CA, USA}
\affil[7]{Center for Immunotherapy and Precision-Immuno-Oncology, Cleveland Clinic, Cleveland, OH, USA}
\affil[8]{National Center for Regenerative Medicine, Cleveland Clinic, Cleveland, OH, USA}
\affil[9]{IBM Quantum, Almaden Research Center, San Jose, CA, USA}
\affil[10]{QuantumBasel, uptownBasel Infinity Corp., Arlesheim, Switzerland}
\affil[11]{Massachusetts General Hospital Cancer Center, Boston, MA, USA}
\affil[12]{Broad Institute of MIT and Harvard, Cambridge, MA, USA}
\affil[13]{Department of Pathology, Massachusetts General Hospital, Boston, MA, USA}
\affil[14]{Department of Pathology, Harvard Medical School, Boston, MA, USA} 
\affil[15]{Algorithmiq Ltd, Helsinki, Finland}
\affil[16]{IBM Research, IBM Research India, India}
\affil[17]{IBM Quantum, IBM Research Europe, Dublin, Ireland}
\affil[18]{IBM Quantum, IBM Thomas J Watson Research Center, Yorktown Heights, NY, USA}
\affil[19]{IBM Quantum, IBM Research Europe, Zurich, Switzerland}

\date{\today}

\maketitle

\begin{abstract}
In recent years, there has been tremendous progress in the development of quantum computing hardware, algorithms and services leading to the expectation that in the near future quantum computers will be capable of performing simulations for natural science applications, operations research, and machine learning at scales mostly inaccessible to classical computers. Whereas the impact of quantum computing has already started to be recognized in fields such as cryptanalysis, natural science simulations, and optimization among others, very little is known about the full potential of quantum computing simulations and machine learning in the realm of healthcare and life science (HCLS). 
Herein, we discuss the transformational changes we expect from the use of quantum computation for HCLS research, more specifically in the field of cell-centric therapeutics. Moreover, we identify and elaborate open problems in cell engineering, tissue modeling, perturbation modeling, and bio-topology while discussing candidate quantum algorithms for research on these topics and their potential advantages over classical computational approaches. 

\end{abstract}

\section{Introduction}

The history of computing is a story of remarkable achievements that continue to transform almost every aspect of our society. 
From the invention of the von Neumann architecture to the emergence of Moore’s law to the recent rise of artificial intelligence (AI), computing has enabled unprecedented advances in natural science, engineering, and medicine~\cite{stafford2010science}.
However, as we approach the physical limits of classical computing, we face new challenges and opportunities that require a paradigm shift in how we process and manipulate information. 
Quantum computing can provide such a paradigm shift. 
 
By harnessing the power of quantum physics, quantum computers can potentially perform some tasks exponentially faster than classical computers and solve problems that are practically intractable for classical computers, such as simulating quantum mechanical systems~\cite{boghosian1998simulating} or decrypting contemporary cryptography~\cite{shor1994algorithms}.
With their exponential rise in computational power (each additional qubit doubles the quantum state space), quantum computers may enable unprecedented transformative potential in the next decades.
However to effectively steer quantum algorithm development and avoid "reinventing the wheel," application domains with vast untapped potential for novel computational approaches need to be identified. 
We believe that, similar to high-energy physics~\cite{dimeglio2023quantum}, healthcare~\cite{floether_2023} and drug discovery \cite{cao2018potential} are prime examples of areas that could see a tremendous impact from quantum-classical computational workflows because they require accurate and reliable simulations of complex systems (for example, molecules, proteins, and cells) or necessitate learning complex behaviors from limited experimental data. In the following, we identify areas in healthcare and life sciences (HCLS) that have seen great advances in the recent years and in which  we believe significant benefits from quantum computing applications will be possible.  

\subsection*{Current technological shift in healthcare and life sciences}
In the past decade, technological advancements have turned biological discovery into an information-rich, quantitative science. From super-resolution microscopy techniques that image macromolecules with nanometer resolution \cite{schermelleh2019super} to organoid technologies that mimic human organs \cite{clevers2016modeling} to spatial single-cell methods able to generate three-dimensional molecular maps of whole tissues \cite{kuett2022three}, new ways of interrogating biological systems across all scales of organization have emerged. These technologies have fueled ambitious efforts, such as creating a human cell atlas, \textit{i.e.} a detailed map of all individual cells in the human body \cite{rozenblatt2017human}, and have transformed how we explore fundamental questions in health and disease. Although most outcomes are still descriptive, studies that show promise in identifying patterns with clinical significance are appearing across a variety of diseases, including cancer \cite{patel2014single, wagner2019single, tirosh2016dissecting}, cardiovascular disease \cite{paik2020single}, and diabetes \cite{segerstolpe2016single}. As a result of such technological advancements, one of the emerging paradigms is the ability to engineer cells to carry out therapeutic functions \cite{lim2010designing,fischbach2013cell}. Reprogramming immune cells has been proven to be successful in treating hematological cancers \cite{WALimScienceReview2022,june2018chimeric,weber2020emerging,hong2020engineering,daniels2022decoding} and the effort has recently been extended to treating solid tumors \cite{reinhard2020rna,de2016cancer,kranz2016systemic} and other diseases \cite{mackensen2022anti,rurik2022car,amor2020senolytic,raffin2020treg}, while also taking advantage of the most recent technologies such as mRNA delivery \cite{pardi2018mrna,pardi2020recent}.

On the computational front, AI (herein defined as intelligent software automating routine labor, understanding and/or recognizing images, text patterns, etc.) and machine learning (ML) (herein defined as the set of algorithms and the mathematical and statistical methods allowing the computer to learn from data) have accelerated discovery in HCLS. 
The use of AI and ML have revolutionized several fields favoring the development and implementation of novel methodologies often based on data-driven approaches. One prominent example of the data-driven solutions provided by AI is in the field of structural biology, where the longstanding problem of predicting the three-dimensional (3D) structure of a protein given a sequence \cite{dill2012protein} has seen significant improvement via transformer-based architectures \cite{jumper2021highly}. 
This work has had a profound effect on the field of synthetic biology by showcasing the potential for using data-driven approaches based on ML methods to solve scientific problems. For instance, in the last few years, novel ML architectures have been developed to generate large protein complexes \cite{shin2021protein,humphreys2021computed,wicky2022hallucinating,bryant2022predicting} and design \textit{de novo} proteins and enzymes \cite{anishchenko2021novo,dauparas2022robust,wang2022scaffolding,mazurenko2019machine,yang2019machine}. 

AI/ML models have been used successfully to predict the effects of noncoding variants \cite{zhou2015predicting} and reach human-level performance in automated whole-cell segmentation, a task that traditionally involved hours of manual processing \cite{greenwald2022whole}. Additionally, several AI/ML approaches have shown great promise in improved disease diagnosis or prognosis. Deep convolutional neural networks (CNNs) have matched the accuracy of radiologists for predictions of lung cancer risk from CT images \cite{ardila2019end}, have outperformed human dermatologists in classification of skin lesions \cite{esteva2017dermatologist}, and have exceeded the performance of established models for breast cancer risk discrimination \cite{yala2019deep}. As AI is rapidly moving towards foundation model-based learning \cite{bommasani2021opportunities}, generalist multi-task medical AI foundation models are emerging \cite{moor2023foundation}. Although the progress in applying AI models to biological data has been impressive, important limitations that hinder their applications to the clinic still persist \cite{kelly2019key}. While some limitations are related to native properties of biological systems, such as their innate complexity and scale \cite{cordier2022biology}, others are associated with shortcomings of AI algorithms, e.g. their inability to learn in data-limited contexts, model overfitting, or learning saturation \cite{ott2022mapping}.

\subsection*{Quantum-enabled healthcare and life science trajectory}
Quantum computing may soon provide researchers with quantum-enabled tools that could expand the limits of computing to unprecedented capabilities, opening up previously unimagined avenues for addressing some of these challenges. Quantum algorithms make use of a radically different computing paradigm that may potentially represent and learn from biological data more efficiently, tackling classically difficult computing problems in healthcare and life sciences. Here, we advocate for the adoption of quantum computing to open up new frontiers for biological research that could enable biomedical discoveries. 
There exist multiple areas for potential significant impact by quantum computing in HCLS that each merit deeper discussions, including biomarker discovery, clinical trial optimization, imaging analysis, and drug protein design and discovery. While naturally there has been a great deal of focus on chemical simulations in biomolecular systems using techniques from quantum simulation \cite{daley2022practical, tacchino2020quantum, altman2021quantum}, here we will focus on applications of quantum machine learning and optimization that have newly realizable potential for healthcare applications owing to recent advancements in quantum hardware and software development. We will elaborate on some of these technological advances in quantum computing from both a hardware and software perspective, with an emphasis on quantum optimization and quantum machine learning.
Importantly, we will present our vision to reimagine healthcare and drug discovery, summarized by \textbf{Quantum Enabled Cell-Centric Therapeutics}, which aims to leverage advancements in single cell and spatial single-cell technologies to create a holistic view of the cellular and metabolic activities in disease tissue to better understand disease dynamics and improve therapeutic design. Here we will highlight four areas of research explorations that address various aspects of a \textit{cell-centric} therapeutic design philosophy and that may serve as a example of how bringing together quantum computing advancements, AI/ML models, and cutting-edge developments in biological research can transform therapeutic discovery and improve healthcare. There is a growing body of work exploring the application of quantum technologies in healthcare, medicine, and the life sciences \cite{floether_griffin_2023}. Hence, we intend for this paper to also serve as a call to both quantum and HCLS researchers to participate in and help shape this vision, devising new biology-inspired quantum algorithms and proof-of-concepts.

\section{Quantum computing state-of-the-art}
Challenges in healthcare and life sciences present opportunities to leverage the unique features of quantum computing to derive novel biological insights to improve patient care.
The scale of today's quantum devices are on the order of tens to hundreds of qubits and remain susceptible to noise \cite{chow_dial_gambetta_2022}. Moreover, the development of error correction techniques used to protect quantum information is still in an early stage \cite{lin2023empirical, google2023suppressing, egan2021fault}. However with recent quantum developments, it may now be the time to begin addressing these HCLS challenges in earnest. 
Qubit counts have increased~\cite{chow_dial_gambetta_2022}, dynamic circuits with mid-circuit measurements have been introduced~\cite{cross2022openqasm}, the fidelities of 1- and 2-qubit gates have improved~\cite{jurcevic2021demonstration}, and the speed of execution of quantum circuits has increased~\cite{wack2021quality}. Advances in error mitigation and error suppression techniques, when coupled with circuit cutting and knitting techniques \cite{piveteau2022circuit, shehab2019noise, patti2022markov}, have enabled researchers to scale up the size of their quantum experiments. 
As an example, quantum chemistry simulations using circuit cutting techniques, such as entanglement forging, have enabled users to utilize double the number of qubits \cite{smith2023clifford} while quantum-classical embedding techniques enabled the scaling up to relevant system sizes~\cite{Rossmannek2021, Rossmannek2023}.
These advancements have demonstrated significant progress towards reaching the stage where quantum computers can solve certain meaningful problems faster, cheaper, or more accurately than classical computers alone for selected applications \cite{kim2023scalable, van2022probabilistic, bravyi2022future}. 
This progress has led to the emergence of quantum machine learning (QML), which aims to use quantum algorithms to analyze large datasets more efficiently than classical machine learning algorithms \cite{cerezo2022challenges}. Depending on the specific use case, quantum computing in general, and QML in particular, may enable different types of benefits, in metrics such as accuracy, energy efficiency, input data requirements, and speed. For instance, there is early evidence in specific instances that quantum algorithms applied to electronic health records are better at handling small, noisy data sets and producing acceptable accuracies than classical approaches~\cite{krunic2022quantum}.
At the end of 2022, IBM announced the production of a 433-qubit processor and that a 100-qubit device capable of achieving a depth of 100 will be available in 2024, representing a new testbed for quantum circuits \cite{IBMQuantumCentricSupercomputing}. It is now possible for quantum circuits to be designed for simple drug discovery problems \cite{cao2018potential,Mensa_2023,Barkoutsos2021, kiss2022quantum} and executed on real quantum hardware using these new capabilities, such as circuit cutting and error mitigation, to complement research on advancing the classical and quantum algorithms.

\begin{figure}[H]
    \centering
\includegraphics[width=.95\columnwidth, keepaspectratio]{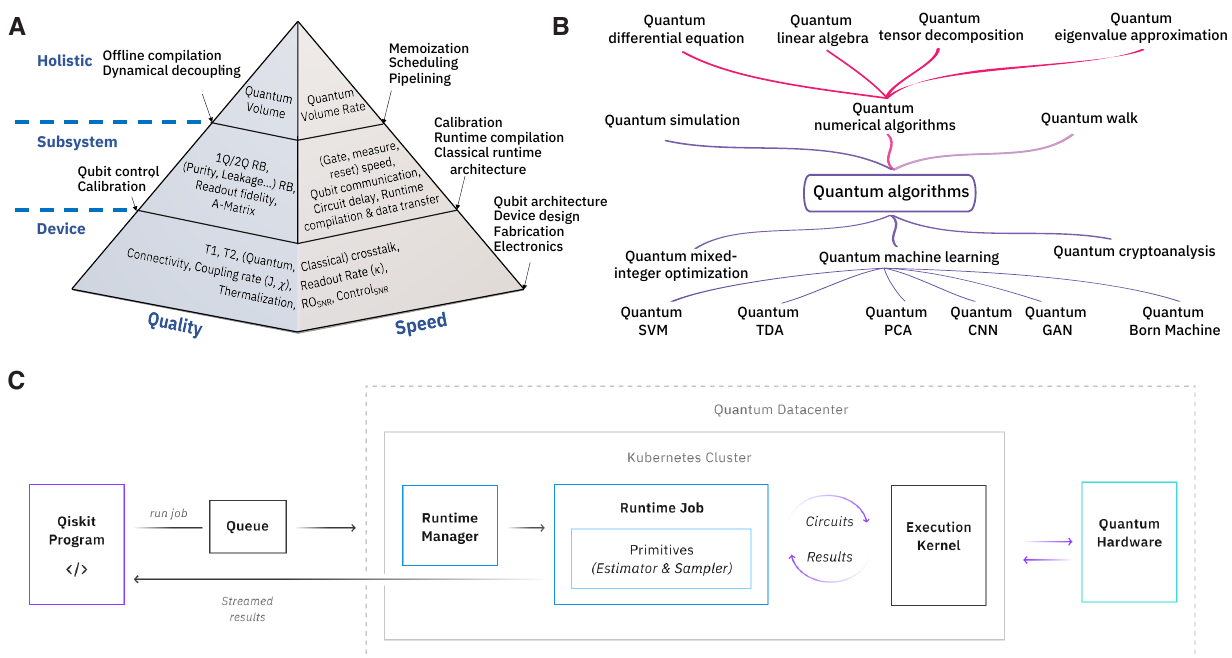}
    \caption{
    \textbf{Quantum computing state-of-the-art.}\\
    \textbf{A}: Reproduced with permission from \cite{wack2021quality}. Different levels of metrics to express the quality of quantum hardware. Along with number of qubits, quantum gate quality is an important quality metric. This is typically expressed in terms of Quantum Volume first introduced in \cite{cross2019validating}. Another quality metric is the speed of execution of quantum circuit layers which is expressed as CLOPS first introduced in \cite{wack2021quality}.  
    \textbf{B}: A mind-map of representative quantum algorithms.
    \textbf{C}: Qiskit Runtime (accessed on May 10, 2023 \cite{IBMQuantumRuntime}) allows a server-less quantum program execution environment. 
    }
    \label{fig:quantum_overview}
\end{figure}

\subsection{Quantum hardware}
Like bits for classical computers, qubits are the basic unit of quantum computation. While classical bits can only be in a state of $0$ or $1$, qubits can exist in a superposition of both states, meaning that they can represent multiple states simultaneously. Additionally, qubits can be entangled, which means that their states are correlated in a way without classical equivalent. Finally, qubits are measured probabilistically and one can measure the interference of their probabilities. These three properties come from the principles of quantum mechanics \cite{griffiths2018introduction}.
There are several approaches to building qubits, including neutral atom qubits \cite{levine2019parallel}, spin qubits \cite{burkard2021semiconductor}, topological qubits \cite{conlon2019error}, trapped-ion qubits, and superconducting qubits, each with its advantages and challenges as assessed by different quantum computing metrics (Fig. \ref{fig:quantum_overview}A). 

While much of the quantum algorithms and applications that we will discuss here are platform-agnostic, we will focus the majority of our discussion on one of the most widely used qubit technologies, superconducting qubits. These qubits rely on the employment of superconducting Josephson's junctions, which leverage the properties of superconducting materials to create and manipulate the two level systems required for quantum computation.
These superconducting qubits are made using superconducting circuits \cite{bravyi2022future, martinis2021optimal, chamberland2022building} that are required to operate at extremely low temperatures, typically just a few degrees above absolute zero, in order to take advantage of the unique property of such materials, which can conduct electricity with no resistance Superconducting qubits are promising candidates for building quantum computers because they can be easily fabricated using standard semiconductor fabrication techniques and have demonstrated long coherence times.

\subsection{Near-term vs. fault-tolerant quantum computing}
Near-term quantum computing refers to the current state of quantum computing technology, where small-scale quantum processors are available for research and development purposes. These devices can perform simple quantum algorithms with limited numbers of qubits.  However, these devices are prone to errors, and must be carefully calibrated and controlled to produce reliable results. Fault-tolerant quantum computing, on the other hand, refers to the theoretical possibility of building large-scale quantum processors that can operate reliably in the presence of noise and errors. These devices would be able to run complex quantum algorithms with many qubits, and would be a major breakthrough in the development of practical quantum computing technology.

The key difference between near-term and fault-tolerant quantum computing (FTQC) is how errors are handled. FTQC uses quantum error correction (QEC) and near-term quantum computing uses error mitigation that trades circuit executions for reduced impact of noise for certain measurements such as observables. In earlier devices, error mitigation \cite{cai2022quantum, kim2023recent} is limited, and the devices are only capable of running small-scale quantum algorithms that can tolerate some errors. Moreover, application of different circuit compilation \cite{madden2022best} and decomposition techniques \cite{piveteau2022circuit} are relatively explicit in near-term quantum computing. However, evidence of quantum utility through quantum error mitigation have started emerging for nontrivial problems \cite{kim2023evidence}. Both the increase in the number of qubits and improved gate quality have contributed to the progress. In fault-tolerant devices \cite{gottesman2022opportunities}, error correction is much more robust, and it requires error rates that are not yet readily achievable, much less at the scale that is needed and hence QEC is just not practical soon. Therefore, advances both in error correction and error mitigation are needed to create a continuous path to quantum advantage. Another important difference is the number of qubits that can be used. Near-term devices typically have on the order of hundreds of qubits, numbers and qubit quality will continue to scale in the near future, where error-corrected devices may require thousands or even millions of qubits \cite{gidney2021factor}. Thus, fault-tolerant quantum computing will require the development of new technologies for qubit fabrication, control, and error correction.

Despite these challenges, researchers are making progress towards fault-tolerant quantum computing, with many promising new technologies and algorithms being developed. However, it is still unclear when or if practical fault-tolerant quantum computers will be built. In the meantime, near-term quantum computing remains an active area of research and development, with many companies and research groups working to build better quantum processors and software tools. These devices are expected to have an impact on fields such as chemistry, optimization, and machine learning.

\subsection{Quantum toolkit and services}
The history of quantum software toolkits in recent years has been characterized by rapid growth and development, as researchers and companies  worked to create more advanced and user-friendly tools for quantum programming. In 2017, IBM released Qiskit, an open-source quantum software toolkit that quickly became one of the most widely used tools in the field \cite{cross2018ibm}. Qiskit runs on IBM Quantum Experience, the first ever cloud-based quantum computing service. Qiskit provides users with a comprehensive set of tools for designing and optimizing quantum circuits, as well as simulating and running quantum programs on real quantum hardware. Other toolkits include Forest \cite{computing2019pyquil}, PennyLane \cite{bergholm2018pennylane}, CirQ \cite{hancockcirq}, Braket \cite{gonzalez2021cloud}, and so on. 
Well-developed quantum software toolkits, like Qiskit, tend to have complex and elaborate modules for different domains of applications. 
Qiskit provides a server-less model to off-load quantum computes to IBM Quantum cloud through Qiskit Runtime as shown in Figure~\ref{fig:quantum_overview}C.
Toolkits, like Qiskit, are essential for enabling researchers and developers to explore the potential of quantum computing, and to build the next generation of quantum applications, such as Algorithmiq’s Aurora~\cite{algorithmiqAurora}, a package for advanced quantum chemistry simulations for drug design and discovery. As these quantum software toolkits continue to evolve and improve, we will see new tools, features, and applications brought online to unlock new areas of HCLS research that are difficult to address using classical computing technologies.

\subsection{Quantum hardware roadmap}
Quantum computing is expected to go through revolutionary changes in the coming years, including expected near term advancements in architectural primitives like modular chip design, classically- and quantumly-connected modules, etc. A combination of larger number of qubits, higher quality quantum gates, and faster clock cycles will allow users to solve significantly larger problems that are common in healthcare and life science.  At the kernel level, threaded primitives, dynamics circuit, built-in error mitigation are also expected to be available at similar timeframe. These will allow an algorithm developer to take advantage of advanced algorithm development techniques from classical computing and expand them into quantum algorithms. Quantum serverless architecture along with advanced circuit decomposition and compilation techniques will enable users to scale up their problem size rapidly. In addition to IBM's announced production of a 433-qubit processor in 2022 and a 100-qubit device capable of achieving a depth of 100 available in 2024~\cite{IBMQuantumCentricSupercomputing}, IBM has targeted the development  of a 100,000 qubit quantum system by 2033~\cite{IBM100000}. In summary, we are entering a new regime in which quantum computers can be used to study and gain insights for important scientific problems, where any non-trivial result will not be easily certifiable with classical alternatives. Deeper collaboration with domain experts is needed to identify the smallest practical problem at that scale. 
Overall, recent developments in quantum computing hardware are driving progress towards the realization of practical quantum computers and empowering quantum algorithms to achieve heretofore unattainable results on actual quantum hardware, and we will provide, as an example, how these quantum advancements can be applied towards a cell-centric therapeutic discovery process.
\\

\subsection{Quantum algorithms}
The history of quantum algorithm development began in the early 1980s when Richard Feynman and Yuri Manin suggested that quantum computers could solve problems that classical computers could not \cite{feynman1985quantum,Manin2023-fz}. In 1994, Peter Shor developed a quantum algorithm for factoring large numbers, which is believed to be intractable for classical computers \cite{shor1994algorithms}. This algorithm is the basis for much of modern cryptography. In 1996, Lov Grover developed a quantum algorithm for searching unsorted databases, which can provide a quadratic speedup compared to classical algorithms \cite{grover1996fast}. In 1997, Seth Lloyd developed a quantum algorithm for simulating quantum systems, which has potential applications in fields such as chemistry and materials science \cite{lloyd1996universal}.
Since then, researchers developed a variety of other quantum algorithms (Fig.~\ref{fig:quantum_overview}B) for tasks such as solving linear equations \cite{lloyd2010quantum}, classification \cite{havlivcek2019supervised},  optimizing functions \cite{farhi2014quantum}, and simulating physical systems \cite{peruzzo2014variational, motta2022emerging}. These algorithms have been further specialized and optimized for different areas of applications including HCLS \cite{cao2018potential, krunic2022quantum}.
\\

\subsubsection{Quantum simulation} 
Quantum algorithms for chemistry simulation are a promising new approach to studying the behavior of atoms and molecules \cite{mcardle2020quantum}. These algorithms exploit the power of quantum computers to simulate the quantum mechanical effects that are essential for understanding the properties of matter. One of the most important quantum algorithms for chemistry simulation is the variational quantum eigensolver (VQE) algorithm \cite{peruzzo2014variational}. VQE can be used to calculate the ground state energy of a molecule, which is a fundamental property that determines its stability and reactivity. The VQE algorithm has been used to simulate the properties of a wide variety of molecules, including water \cite{nam2020ground}, methane \cite{gokhale2020optimization}, deuteron \cite{dumitrescu2018cloud, shehab2019toward}, and large organic molecules such as butyronitrile~\cite{Rossmannek2023} and ferrocence~\cite{fitzpatrick2022selfconsistent}.

Quantum algorithms are also being developed to simulate excited states of quantum systems. Excited state simulations are needed for understanding the behavior of many physical and chemical systems, including electronic and magnetic properties of materials. While VQEs have been used to find these excited state energies of a quantum system by optimizing a variational ansatz wave function \cite{higgott2019variational}, quantum subspace expansion (QSE) and the quantum equation-of-motion (qEOM) algorithms~\cite{OllitraultqEOM2020} have also been developed to perform excited state simulations. \cite{motta_quantum_2023} These methods simulate the excited states of a quantum system by expanding the wave function in a basis of excited states \cite{mcclean2017hybrid}.
 The quantum Lanczos algorithm has also been found useful for excited state simulations \cite{kirby2022exact} by efficiently computing the eigenvalues and eigenvectors of a Hamiltonian matrix, which can be used to determine the excited states of a quantum system. In addition to these specific algorithms, there is growing interest in developing general-purpose quantum algorithms for excited state simulations. These algorithms would be able to simulate a range of quantum systems and phenomena, including the excited states of many-body systems. The development of quantum algorithms for ground and excited state simulations is still in its early stages, but researchers are optimistic about the potential impact of these algorithms on the field.

\subsubsection{Quantum operations research}
Quantum algorithms for operations research is a rapidly evolving area of research that focuses on applying quantum computing to solve problems in operations research. Quantum Monte Carlo algorithms can simulate the behavior of quantum systems, which is useful for solving problems in optimization, finance, and other fields \cite{patti2022markov}. Quantum algorithms for mixed-integer programming problems are able to exploit problem structure to maximize fractional Grover speedup \cite{gambella2020multiblock}.
Quantum walk algorithms may have the potential for quantum advantage over classical Monte Carlo or random walk algorithms \cite{szegedy2004quantum, wocjan2021szegedy, childs2009universal}. Additionally, a large number of optimization problems may be represented as quadratic unconstrained binary optimization (QUBO) problems which allows quantum approximate optimization algorithm (QAOA) to create the potential of maximizing fractional Grover speedup \cite{farhi2014quantum, stkechly2023connecting}.

\subsubsection{Quantum machine learning}\label{section:qml}
Quantum machine learning (QML) is an evolving field that combines the power of quantum computing with the techniques of machine learning~\cite{biamonte2017quantum, schuld2015introduction, cerezo2022challenges}. QML algorithms have the potential to significantly improve the accuracy and efficiency of machine learning models, particularly for large datasets. One key advantage of QML is the ability to perform certain computations exponentially faster and more efficiently than classical computers~\cite{havlivcek2019supervised, liu2021rigorous}. Another advantage is the potential for QML to solve problems that are not easily solved with classical techniques. QML algorithms can be used for a variety of tasks, including classification, regression, clustering, and dimensionality reduction. Some popular QML algorithms include the quantum support vector machine~\cite{Rebentrost_2014}, quantum k-means clustering~\cite{Kavitha2022}, and quantum neural networks~\cite{abbas2021power}. While QML is still in its early stages, researchers and industry experts are optimistic about its potential impact on fields such as drug discovery, finance, and materials science. However, one major challenge of QML is the need for large-scale, fault-tolerant quantum computers, which are not yet available. More importantly, the quantum amenability to practical datasets for machine learning is still not well understood. Nevertheless, quantum machine learning is an exciting area of research with promising possibilities for advancing the field of machine learning.

\vspace{1cm}
\noindent\textbf{Quantum Support Vector Machine} The quantum support vector machine (QSVM) is an emerging approach that combines principles from quantum computing and classical machine learning to enhance the capabilities of support vector machines (SVMs) \cite{havlivcek2019supervised}. SVMs are powerful algorithms used for classification and regression tasks, but they face limitations when dealing with large and complex datasets. QSVMs aim to overcome these limitations by leveraging quantum feature maps that encodes information into a Hilbert space. Fundamentally, the QSVM algorithm employs a quantum kernel function that measures the similarity between quantum feature vectors as the inner product between corresponding density matrices. These feature vectors encode information about the input data and are represented as quantum states. The kernel function can be computed by running the quantum circuit for feature maps corresponding to one of the inputs followed by the inverse circuit for the other quantum circuit and finally measuring in standard basis. One can then optimize the dual on a classical machine based on the kernel function computed above on quantum circuit. While the theoretical foundation of quantum feature maps is still an area of active research, asymptotic quantum speedup has been demonstrated for a certain quantum feature map \cite{liu2021rigorous}, and empirical quantum advantage has been demonstrated for electronic health record data in a few very specific settings \cite{krunic2022quantum}.  

\vspace{.25cm}
\noindent\textbf{Quantum Neural Networks} The convergence of quantum information science and machine learning has given rise to a novel and promising paradigm known as quantum neural networks (QNNs) \cite{biamonte2017quantum, cerezo2022challenges}. QNNs combine the principles of quantum mechanics and classical neural networks. Like their classical counterparts, QNNs consist of interconnected nodes, or "neurons", that process and transmit information. However, unlike classical neural networks that rely on classical bits, QNNs employ qubits as their basic units of information.  The architecture of a QNN comprises three key components: the input layer, hidden layers, and output layer. Each layer consists of qubits and quantum gates. The input layer encodes the input data, which is then processed through the hidden layers via quantum gates. Finally, the output layer produces the desired output based on the learned patterns. Training QNNs involves adjusting the parameters of the quantum gates to optimize the network's performance. Quantum algorithms, such as quantum gradient descent and quantum variational algorithms, play a crucial role in the training process. Quantum gradient descent adapts the parameters of the gates by minimizing the loss function, while variational algorithms optimize the parameters by leveraging quantum optimization techniques. It has been shown that a class of quantum neural networks is able to achieve a  better effective dimension, which is a robust capacity measure, than comparable feed-forward networks and can be trained faster \cite{abbas2021power}.

\vspace{.25cm}
\noindent\textbf{Quantum Topological Data Analysis} Topological data analysis (TDA), introduced in early 2000s \cite{carlssonTDA, edelsbrunnerTDA} is a novel data science method that combines tools from algebraic topology and computational geometry to study and analyze the shape of the data to reveal hidden structures and patterns while gaining insights independent of noise. Some successful applications of TDA include areas such as medicine, biology, image analysis, network analysis, and multi-variate time series analysis \cite{TDAbiologyfolding, TDAmedicinearrhythmia, TDAimageneural, TDAtimeseries}. In particular, TDA can efficiently extract higher dimensional features from a noisy data set. While TDA found many applications in different research areas, including healthcare and life sciences~\cite{Dey2022,Masoomy2021,Loughrey2021,Rabadan2020,Camara2017,Nicolau2011}, it is known that the computational cost increases exponentially in the number of data points or in the dimension of the topological features targeted. The first quantum algorithm for TDA offered an exponential speed up targeting the regime where classical TDA struggles, in particular higher dimensional features \cite{LGZalgorithm}. Under certain constraints, this algorithm utilizes some standard quantum protocols such as multi-targeted Grover's search algorithm and quantum phase estimation (QPE). This created a surge of attention to develop improved quantum algorithms for TDA \cite{ReviewQTDA, QTDAtowardsadv, QTDAhayakawa, schmidhuber2022complexitytheoretic, QTDAlindepth, QTDAlinexp}, providing both theoretical and experimental framework for researchers to analyze higher-dimensional data using quantum computing.

\vspace{.25cm}
\noindent\textbf{Quantum Cumulant Calculation} Cumulant calculation is a mathematical technique that can be used to analyze and understand complex, high-dimensional, and noisy data sets including healthcare data, and has been used in conjunction with TDA to better capture high-dimensional relationships. By using cumulant-based analyses one can identify patterns and relationships within the data, which are often challenging to extract using traditional statistical methods. 
Cumulants provide a way to identify redescriptions \cite{parida_l_redescription_2005}, which can be used to identify and generate logical relationships among variates that may indicate underlying biological processes, and whose connectivities provide information about distinct pathways to disease \cite{platt_character2016,karisani_topology_2022}.  The idea of redescriptions is that patients who develop conditions (e.g. atherosclerosis) tend to share clusters of other comorbidities (e.g. hypertension) and can be captured in groups according to their diagnoses of these conditions.  
Often, healthcare data are strongly correlated and therefore higher-order moments tend to include the impact of such lower order correlations. These higher-order associations may reveal specific biological pathways driving disease and can lead to distinguishing multiple distinct pathways captured by edges in a network formed by multi-omics variables~\cite{platt_epidemiological_2022, bose_cuna_2021}. It is possible to exclude potential spurious lower-order correlations by computing cumulants of higher-order moments involving products of collections of variates since cumulants vanish if the products of collections of variates partition into independent subsets~\cite{percus_correlation_1975}. In order to study all the moments or the cumulants of a sequence of random variables it is usual to consider the moment as well as the cumulant generating functions. These are functions of $n$ indeterminates where $n$ is the number of random variables in question. These generating functions, which contain all the information about the moments and cumulants, can be quite complicated and can be challenging to compute. An analogous situation occurs in statistical physics. The partition function in various Ising models is an exponential sized sum over all configurations where the indeterminates correspond to magnetization variables. Computing the partition function is equally challenging as computations of moments or cumulants because of the exponential sized sums. Various approaches have been suggested for computing good approximations. In particular, taking all the magentization variables to be equal one obtains a polynomial in one variable. The study of the zeros of this univariate polynomial (via the Lee-Young theorem~\cite{lee_statistical_1952}) gives important information about the macroscopic behavior of the corresponding physical system -- such as phase transition. Efficient quantum circuits has been developed for computing the complex zeros of the partition function. An analogous quantum approach towards the computation of moments and cumulants will be very interesting.

\vspace{.25cm}
\noindent\textbf{Quantum Network Medicine} The emerging field of network medicine~\cite{Barabasi2011} applies network science approaches to investigate the molecular complexity of a particular disease, as well as the molecular relationships among apparently distinct phenotypes, integrating information from relevant Omics databases. The modern era has seen an exceptional growth in molecular interaction data such as molecular networks, including protein interaction networks, whose nodes are proteins that are linked to each other by physical interactions; metabolic networks, whose nodes are metabolites that are linked if they participate in the same biochemical reactions; and regulatory networks, whose directed links typically represent regulatory relationships between a transcription factor and a gene. However, such networks are known to be vastly incomplete, with large proportions of the true interactions being yet unknown~\cite{Luck2020}. Moreover, if we are to efficiently search for new drugs and drug combinations or pathogenic interactions within and between cells, there is a pressing need for computational methods that can access the immense molecular space until now largely unexplored. Quantum computing may be a key ingredient in enabling the full potential of network medicine. Recently, it has been proposed to combine network medicine and quantum algorithms in a novel research field, quantum network medicine, to lay the foundations of a new era of disease prevention and drug design~\cite{maniscalco2022quantum}. A successful example of the potential of this field is the recent demonstration of a link prediction algorithm based on continuous-time quantum walks~\cite{Goldsmith2023}. This algorithm has also been successfully adapted to identify disease modules and new disease pathways.

\FloatBarrier
\section{Quantum-enabled cell-centric therapeutics}
Therapeutic design and discovery has traditionally focused on drug-target identification and interaction optimization, and there exist many classical approaches to perform such analyses as well as clear quantum applicable algorithms. This target-centric approach has been the dominant paradigm in therapeutic design and led to the successful approval of many novel therapeutics (e.g., small molecule inhibitors, chemotherapeutic, antibody therapies) across a multitude of diseases. However, the cost in research \& development per new approved drug has been doubling $\sim$ nine years since the 1950s~\cite{scannell2012diagnosing} and, for many diseases (e.g., rare diseases or particularly aggressive cancers like pancreatic), effective therapies are still far away. 
While typically still considered the gold standard, the target-centric approach to drug discovery appears to be falling short on delivering significant numbers of therapeutic advances~\cite{moffat2017opportunities}, 
which may be attributable to the complexity and uncertainty in target validation~\cite{sams2005target}.
The validity of target-centric approaches may be reaching the point of diminishing returns as evidenced by the observation that many anticancer drugs in clinical trials exert their efficacy not through their ostensible mechanism of action but via some off-target cytotoxicity~\cite{lin2019off}. Moreover, medicine itself is transforming. As we progress towards precision medicine, treatments at the level of an individual, and even better proactive interventions to keep people healthy, are needed. Today, many standardized therapies fail to achieve their intended outcomes, a case in point being that less than half of cancer patients respond to immunotherapy~\cite{sharma2017primary}, thus necessitating more tailored methods~\cite{abbott2021quantum}.

\begin{figure}[ht]
    \centering
\includegraphics[width=.99\columnwidth, keepaspectratio]{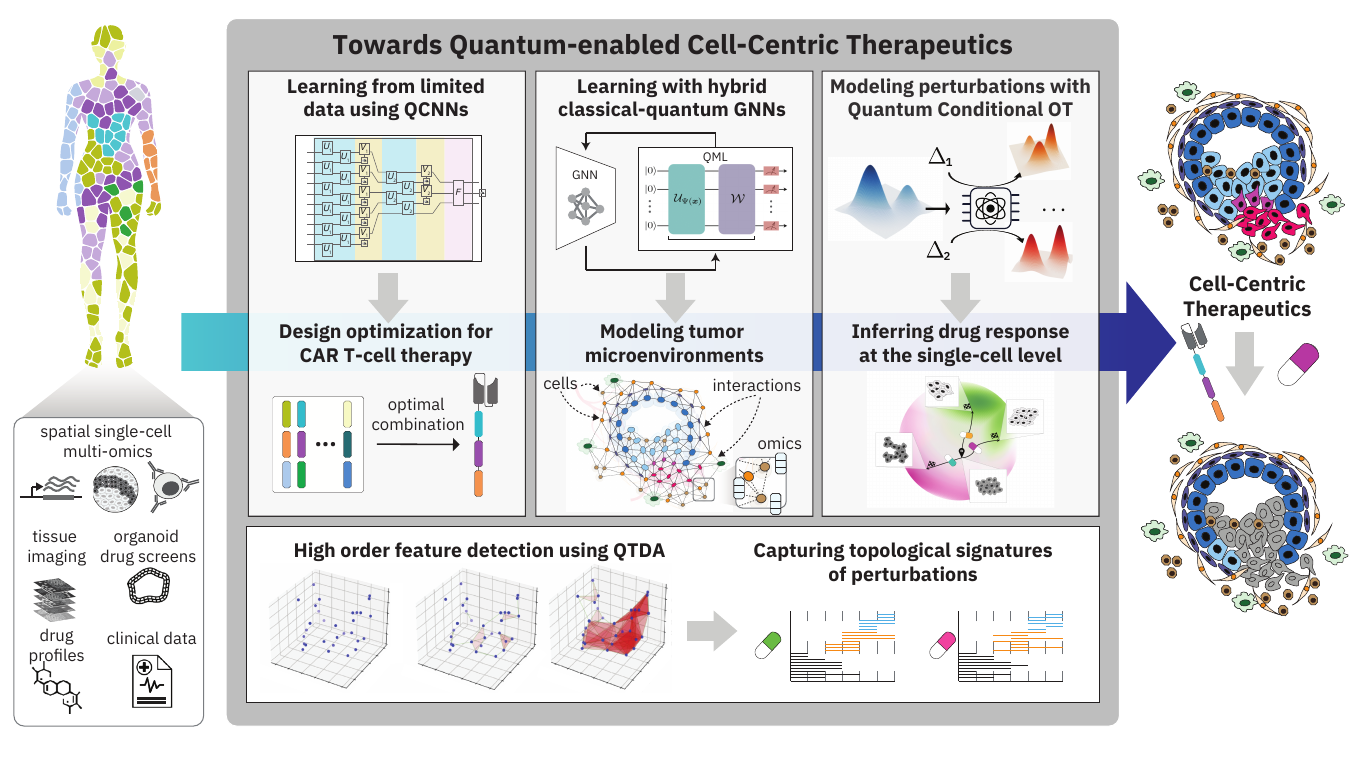}
    \caption{\textbf{Overview of quantum-enabled cell-centric therapeutics.}\\
    Spatiotemporal single-cell, cell-line, imaging, drug profile, and clinical data are analyzed with four quantum computing technologies to capture varying aspects of cellular behavior. These technologies include: (top left) QCNNs to learn optimal CAR T-cell intracellular signalling domain design from limited experiment data; (center) hybrid classical-quantum GNNs to model tumor microenvironments from single cell spatial data; (top right) single cell perturbation response using Quantum Conditional OT; and (bottom) QTDA to identify topological signatures of single cell perturbation response. Using insights from each research area and by combining the efforts between them, we aim to provide treatment options that optimize the cellular context to improve therapeutic response.\\
    }
    \label{fig:overview}
\end{figure}

With the advent of single-cell and spatial single-cell technologies there is now an opportunity to more precisely understand the interactions between disease cells, microenvironment, and therapeutics to accelerate \textit{cell-centric} therapeutic approaches and precision medicine. These technologies provide a detailed accounting of the activities occurring within each cell and of how those cells are interacting in a tissue. In addition, they enable the capture of a holistic perspective of the cellular and metabolic activities of malfunctioning tissue with single-cell precision, allowing the study of disease dynamics in highly heterogeneous tissues with unprecedented spatiotemporal resolution. The trajectory of these technological advancements points towards a transformation of therapeutic design. This is made more relevant as we seek to characterize complex diseases with heterogeneous disease pathologies that cannot be described through single-protein effects, and where a conventional target-centric approach cannot be successful without the involvement of serendipity~\cite{hargrave2012serendipity}.

By using the data obtained from these new technologies to examine the cellular state with respect to its cellular context and in combination with the potential of quantum computing to complement classical computing techniques, a cell-centric therapeutic design can be realized. Under this regime, the goal is to identify therapeutics, biologics, or other interventions that modify the cellular disease ecosystem to make it more responsive to therapy or shift it to a quiescent or even dying state. This is achieved by modeling the disease  microenvironment at single-cell resolution to understand the cellular interactions, feature space, and the needed changes that shift the environment from a non-responsive to a responsive state. 

Though classical AI methods have shown themselves to be critical in analysing single cell and spatial single cell data, there remain challenges where quantum computing approaches may offer significant advantages. Single cell omics data is often high dimensional and with high sparsity where most genes are not altered or expressed in any given single cell. These cells are typically not labeled and thus semi-supervised approaches are needed to learn from labeled and unlabeled single cell data. These challenges in single cell analysis are being addressed with breakthrough advancements in transfer learning and transformer models~\cite{Chen2023, Theodoris2023, Cui2023, Hao2023}. Yet there remains important limitations with these classical computing technologies, including requiring large amount of training data, poor explainability or interpretability, limitations in capturing global contextual signals, inability to control attention, and quadratic space and time complexity. 
These challenges are exacerbated when data sets from related techniques are combined, for example from flow cytometry and single-cell sequencing~\cite{andreyev2020integration}; classical methods struggle with leveraging such heterogeneous data to effectively classify cells based on their physical and biochemical characteristics.

A cell-centric approach to understand perturbation response and disease behaviors presents an unique opportunity for multiple quantum and quantum-classical optimization and machine learning techniques to be brought together to address these challenges and further advance cell-centric therapeutic design strategies. 
In the following, we will describe several avenues of research capturing varying aspects of this cell-centric approach and how they may come together by focusing on a cancer use case (Fig.~\ref{fig:overview}). These include approaches for optimizing CAR T-Cell engineering; representing and analyzing spatial, single-cell data; developing predictive models of single-cell perturbation; and extracting n-th order feature interactions that inform cellular behavior.

\subsection{Cell engineering for immunotherapy}

\begin{figure}[ht]
    \centering
\includegraphics[width=.98\columnwidth, keepaspectratio]{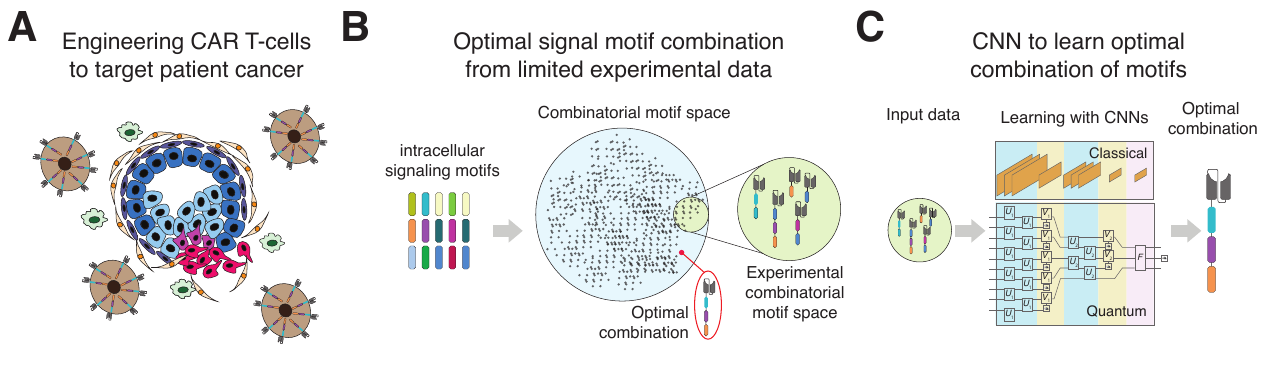}
    \caption{\textbf{Quantum Convolutional Neural Network for cellular engineering.}\\
    \textbf{A}: Cartoon of a Chimeric antigen receptor (CAR) T-cells (brown colored) surrounding a tumor with its microenvironment. CAR T-cells with both the engineered extracellular and intracellular domains of the receptor are shown.  
    \textbf{B}: Schematic representation of the combinatorial motif domain derived from a library of signal motif domains.  These domains can be arranged in up to three positions within the CAR T-cell intracellular region. Each combination gives rise to specific CAR T-cell phenotypes. The combinatorial space (cyan) is typically much larger than the experimentally generated CAR T-cells used in development (green). The optimal motif combination giving rise to optimal CAR T cell phenotype may fall outside of what was experimentally generated (red).
    \textbf{C}: Schematic representation of the learning of the optimal combination of the signalling motifs using convolutional neural networks, either current classical CNNs (top) or quantum CNNs (below). For the learning process, experimental data are used as input data.
    }
    \label{fig:quantum-cell-engineering}
\end{figure}
Cell therapies are powerful new medicines in which a class of human cells are reprogrammed to carry out specific functions such as killing cancer cells (an example is represented schematically in Fig. \ref{fig:quantum-cell-engineering}A). These therapies offer novel approaches that eventually will lead to treatments of other diseases such as autoimmune disorders, inflammation, and neuro-degeneration, by overcoming immunosuppressive tumor microenvironments, reducing toxicities, and preventing antigen escape. The great advantage of cell therapies lies in the possibility of engineering each cell modular component, each characterized by specific features, and building a synthetic molecule or cell with desired functions. A cell's phenotype can be reprogrammed by engineering cells at different scales, whether single point amino acid mutations \cite{guedan2020single}, designed peptides (e.g. antimicrobial peptides) \cite{das2021accelerated}, protein receptors \cite{krause1998antigen,roybal2016engineering,tousley2023co,roybal2016precision}, or cells and tissues \cite{martinez2022scaling,harrison2017assembly,shahbazi2018deconstructing}. The downside of such reprogramming is that the possibility of engineering and combining different cell modular components generates a vast, complex combinatorial design space that is difficult to explore experimentally. For this reason, AI/ML models are powerful tools to address this challenge because such models can learn complex patterns and features from a given dataset and provide predictions on different phenotypes of interest.

Chimeric antigen receptors (CARs) are genetically modified T cell receptors designed to repurpose the phenotypic output of natural T cells. The extracellular domain of CAR T cells is engineered to identify specific tumor-associated antigen and the intracellular domain is engineered to include intracellular motifs that enhance the T cell activation and function (Fig. \ref{fig:quantum-cell-engineering}A). These motifs are part of the co-stimulatory domains and are responsible for the antitumor efficacy of CAR T cells. Combinations of these domains generate CAR T cells with different features, and currently, six CAR T cell therapies for 12 applications are approved by the US Food and Drug Administration (FDA) \cite{labanieh2023car}. Hence, given the proven efficacy of this new technology, the outstanding question is how can we efficiently explore all the possible costimulatory domain combinations to design optimal CAR T cell for a given patient? Since there are many potential combinations of associated sets of signalling and activation motifs for an engineered T cell, novel approaches based on screening pooled CAR signaling domain libraries have been proposed \cite{gordon2022screening,castellanos2022speedingcars}. Specifically, KG Daniels et al.~\cite{daniels2022decoding} seeking to optimize the phenotypic response of CAR T cells, the authors defined a combinatorial library of 13 motifs located in 3 different positions of the receptor intracellular domain (Fig. \ref{fig:quantum-cell-engineering}B). With $\approx2350$ potential combinations, only $\approx250$ were tested experimentally (Fig. \ref{fig:quantum-cell-engineering}B) and a ML model was used to predict the cytotoxicity of the other remaining combinations. The ML algorithm, based on a Convolutional Neural Network (CNN) (\ref{fig:quantum-cell-engineering}C) with long-short term memory (LSTM), was able to reach $\approx70$\% accuracy at predicting CAR T cell phenotype. 

Given the highly data-constrained problem described above, we have identified this study as an instance where quantum neural networks (QNNs) would potentially be able to provide advancements. Adding to the power of classical neural network models, QNNs utilize quantum mechanical effects such as superposition, entanglement, and interference to represent complex relations among data. As such, certain QNN architectures have been shown to have greater expressivity than some of their classical counterparts, allowing them to capture more complex probability distributions than classical models, and additionally indicating that there might be potential speed-ups in training time. In general, identifying problems where QNNs are more advantageous than classical models in training or model accuracy remains an open question with great room for improvement and requires heuristic experimentation with the datasets considered. As such the use of Quantum Convolutional Neural Networks (QCNNs)~\cite{cong2019quantum} (Fig. \ref{fig:quantum-cell-engineering}C) can be employed to improve the aforementioned 70\% accuracy. QCNNs have several useful properties, including the number of variational parameters that scale logarithmically with the number of qubits and the absence of barren plateaus during training that can affect other types of QNNs~\cite{pesah2021absence}. Finally, it has been shown that some quantum ML models, such as QCNNs, can reach low generalization errors even in the case of limited training data, which further motivates the use of QCNNs for this problem domain.

\subsection{Modeling tumor microenvironments with hybrid classical-quantum GNNs }

\begin{figure}[ht]
    \centering
\includegraphics[width=.98\columnwidth, keepaspectratio]{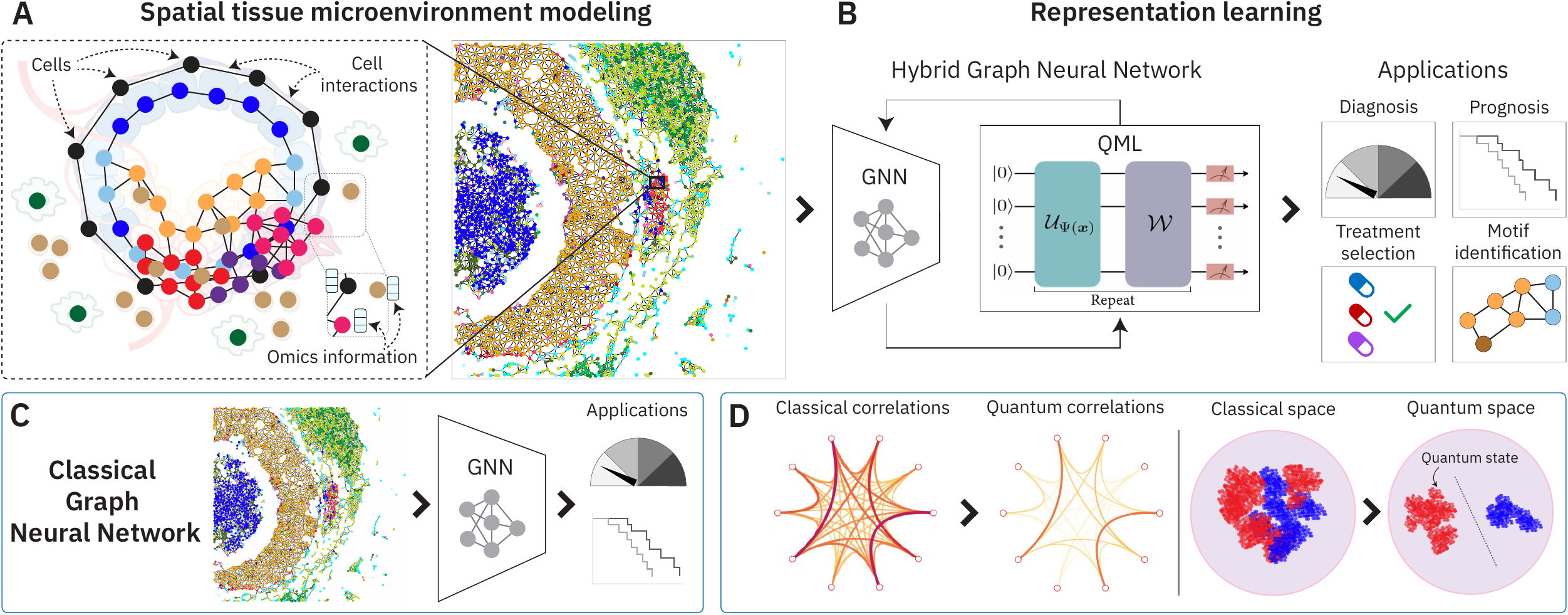}
    \caption{\textbf{Quantum Geometric Machine Learning for spatial tissue modeling.}\\
    \textbf{A}: Spatial distribution of cells in a tissue microenvironment is modeled as a cell-graph, where nodes are cells, edges are cell-to-cell interactions, and  
    nodes encode cell-level multi-omics information.
    \textbf{B}: Learning an effective representation of a cell-graph can address immediate clinical applications (diagnosis, prognosis, treatment selection) as well as discover crucial patterns driving the disease. To this end, we propose a novel hybrid classical-quantum Graph Neural Network (GNN) by integrating the capabilities of GNNs and quantum machine learning.
    \textbf{C}: Though a classical GNN operating on a cell-graph can tackle the aforementioned applications, it poses a number of crucial challenges (e.g., a sub-optimal embedding space, ineffective in addressing $NP$-hard combinatorial tasks) that can benefit from Quantum Computing.
    \textbf{D}: The proposed algorithm extends the classical embedding space to exponentially large qubit Hilbert space and captures hidden correlations among the cells to improve class separability. Showcased on spatial tissue modeling for various downstream applications, this algorithm finds general utility in other graph-structured data.
    }
    \label{fig:spatial_tissue_modeling}
\end{figure}

Spatial single-cell omics are currently revolutionizing and paving way for advancing cancer biology by enabling deep phenotypic profiling of each individual cell within the tumor ecosystem while preserving its topology \cite{lewis2021spatial}. 
From multiplexed imaging to spatial single-cell transcriptomics, such spatial data can be elegantly modeled using cell-graph representations (Fig. \ref{fig:spatial_tissue_modeling}A), where cells are the nodes, including cell-specific information, and edges denote cell-to-cell interactions.
Consequently, Graph Neural Networks (GNNs) have found initial applications in learning on spatial single-cell data (Fig. \ref{fig:spatial_tissue_modeling}C).
Modeling and learning on spatial cell-graphs exhibit several attractive properties: GNNs explicitly learn on cells instead of encoding pixels, and can elegantly integrate single-cell information with disease tissue (e.g. tumor) morphology, topology, and interactions among cells and/or tissue structures \cite{hetzel2021graph, li2022graph}. The cell-graphs bestow an interpretable input space, which enables incorporation of prior domain knowledge from medical experts. 
Furthermore, GNNs are not limited by variations in resolution and can be easily coupled with explainability techniques to provide valuable insights on which cells or cell neighborhoods drive the decision \cite{jaume2021quantifying}. 
Initial attempts to leverage the advantages of GNNs on spatial single-cell data have started to emerge, with applications in cell phenotyping \cite{brbic2022annotation}, learning cell-cell communication \cite{fischer2022modeling, yuan2020gcng} and, recently, modeling tumor microenvironments \cite{wu2022graph}. Yet, the complexity of tumor graphs and the entangled cell neighborhoods lead to suboptimal embedding spaces of GNNs, which in turn struggle with learning clinically meaningful patterns from the data. At the same time, searching for relatively small query subgraphs over large, complex graphs is an $NP$-hard problem where classical computing approaches do not suffice. 

These limitations present an interesting opportunity for quantum computing. Mapping spatial data to the exponentially large Hilbert spaces can potentially solve the sub-optimal embedding of cell neighborhoods, searching for small query sub-graphs within large graphs can be improved, and leveraging the state-space available in quantum can lead to higher predictive ability of QML models. It is also possible for quantum computing to implement biases and symmetries as well as capture hidden correlations more efficiently (Fig.  \ref{fig:spatial_tissue_modeling}D). 
Currently, hybrid quantum-classical solutions are implemented that combine GNNs for data pre-processing with QML algorithms, such as Variational Quantum Classifiers (VQC), Quantum Neural Networks (QNN), and Quantum Support Vector Machines (QSVM), and validated for downstream tasks such as tumor subtyping \cite{brancati2022bracs,bulten2022}  (Fig. \ref{fig:spatial_tissue_modeling}B). 
Quantum versions of GNNs can also be created and researched upon for these inherent graph problems to study the possible advantages quantum can provide as new research directions and advance the state-of-art in the spatial single-cell omics.

\subsection{Inferring single-cell drug perturbations with quantum conditional optimal transport}

\begin{figure}[ht]
    \centering
\includegraphics[width=.95\columnwidth, keepaspectratio]{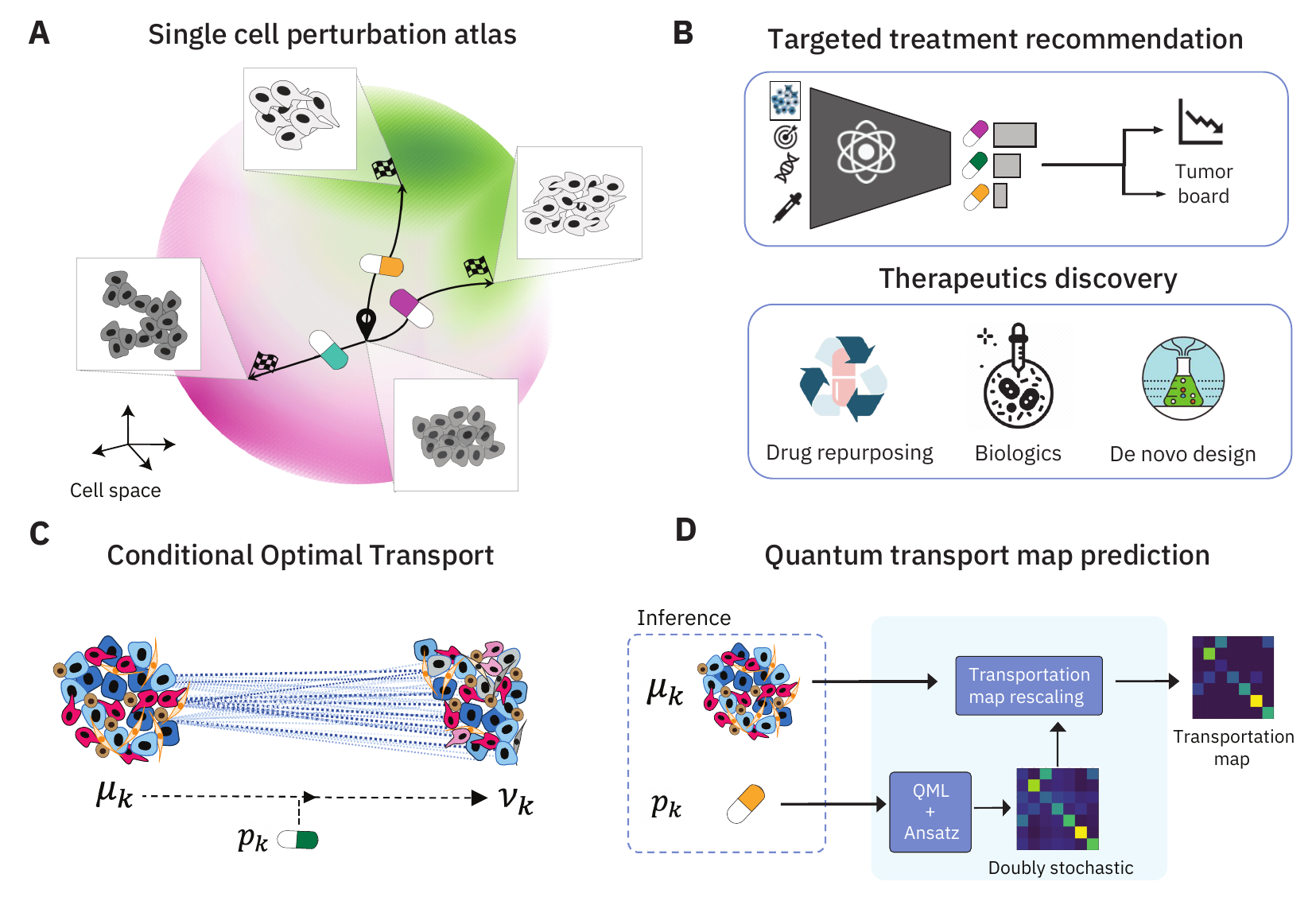}
    \caption{\textbf{Quantum Conditional Optimal Transport for single-cell perturbation modeling.}\\
    \textbf{A}: In a single cell perturbation atlas with a given initial cell state (center), different perturbations (in this case drug administrations) result in different cell states with varying favorability.
    \textbf{B}: Modeling perturbational effects of biological systems with single cell resolution may impact immediate clinical applications (personalized treatment recommendation) as well as discovery of novel therapeutics.
    \textbf{C}: Rooting in optimization and transportation theory, \textit{Conditional} Optimal Transport recently emerged as a parametric approach to estimate transportation plans conditioned on a context variable.
    \textbf{D}: Building upon OT principles, we propose a novel Quantum algorithm for parameterizable transportation map estimation that leverages unitarity as well as Sinkhorn rescaling.
    Showcased on single cell perturbation tasks, this algorithm finds general application in simulating effects of interventions on a given system.
    }
    \label{fig:qondot}
\end{figure}
At the core of finding effective, novel therapeutics lies understanding the tissue response to specific therapeutic interventions, such as drug administrations.  
While cell line perturbation studies have long been successfully used for pre-clinical validation of targeted cancer drug candidates, recent technological developments have allowed for such perturbation studies at the single-cell level~\cite{srivatsan2020massively}.
In these studies, the multiomic states for each individual cell are measured \textit{before} and \textit{after} drug or genetic perturbations resulting in perturbation atlases that also capture the underlying heterogeneity of drug response. 
These highly informative atlases facilitate prediction of perturbation response on the tumour tissue at the resolution of single-cells, in the near future even for entire tumors (Fig.~\ref{fig:qondot}A). 
The overarching long term objective of computational models in this field will be to simulate responses of tumor microenvironments to therapeutic interventions, e.g. drug administrations, and, ultimately, to develop trajectories that capture tumor growth patterns within its native microenvironment.
Early attempts in this direction used autoencoders with linear latent arithmetics and achieved reasonable success at approximating the effect of such interventions~\cite{lotfollahi2019scgen}. 
Notably however, the problem formulation is deeply rooted in optimal transport theory, which in simple intuitive terms can be illustrated by minimization of the earth mover distance when moving one distribution of earth to another~\cite{villani2009optimal}.
Thanks to approximation techniques, such as entropic regularizations~\cite{cuturi2013sinkhorn}, Optimal Transport (OT) recently gained popularity across machine learning applications~\cite{peyre2019computational}.
Contemporary work suggests that the posed task can be approached best with \textit{conditional} OT~\cite{bunne2022supervised} -- which leverages the OT principles, thus benefits from strong theoretical support and explainability while allowing to \textit{condition} the OT on a desired perturbation (Fig.~\ref{fig:qondot}C).

We propose a novel hybrid quantum optimization algorithm for single-cell perturbation modeling that is able to predict transportation maps tailored to individual patients and therapeutic hypotheses. 
While the theoretical support for such a quantum OT algorithm is in development, this formulation exploits some natural links between unitary operators and the structure of the OT maps (Fig.~\ref{fig:qondot}D).
The proposed system is currently validated on several public datasets of single-cell perturbations of different drugs, drug dosages, and CRISPR gene knockouts~\cite{srivatsan2020massively,norman2019exploring,gehring2020highly}; with the objective to achieve a significant performance improvement in capturing therapeutic effects on a tumor level. Such a methodology naturally integrates with the previously discussed activities, in particular the spatial tumor modeling work which facilitates downstream clinical applications, e.g. recommendation of therapeutic interventions, based on the simulated tumors (Fig.~\ref{fig:qondot}B). 
The method is built largely in anticipation of novel single-cell data acquisition techniques that will arise in the near future, such as spatiotemporal perturbation data or combinatorial perturbations. 
Future extensions of this quantum framework will focus on modeling cell system evolution over time, e.g., reconstructing trajectories of cell differentiation.
Remarkably, the devised conditional OT methodology is generic and finds immediate application across a rich set of problems in the healthcare and life sciences -- essentially all scenarios where the state of a system is captured \textit{before} and \textit{after} a certain intervention and where the goal is to understand how the intervention alters the \textit{before} state to a hypothetical \textit{after} state.

\subsection{BioTopology for cellular behavior}
Combining insights from both a spatial cell model and cell perturbation model enabled with quantum would provide valuable insights into how tumors evade therapeutic response. Yet it is known that despite the incredible amount of information contained in imagery and omics data, this information is incomplete. This incompleteness may be due to throughput constraints on technology, constraints on funding, measurement variances between technology platforms, or lacking the technology to measure every relevant molecular entity within a cell at once. Furthermore, cell behaviors and hence disease phenotypes are the result of complex relationships dynamically occurring over time that can not be captured at scale. Hence novel methods are needed to discover those cryptic relationships, likely residing in connections among significant higher order interactions between the dimensions of a given feature space.

\begin{figure}[htb!]
    \centering
\includegraphics[width=.95\columnwidth, keepaspectratio]{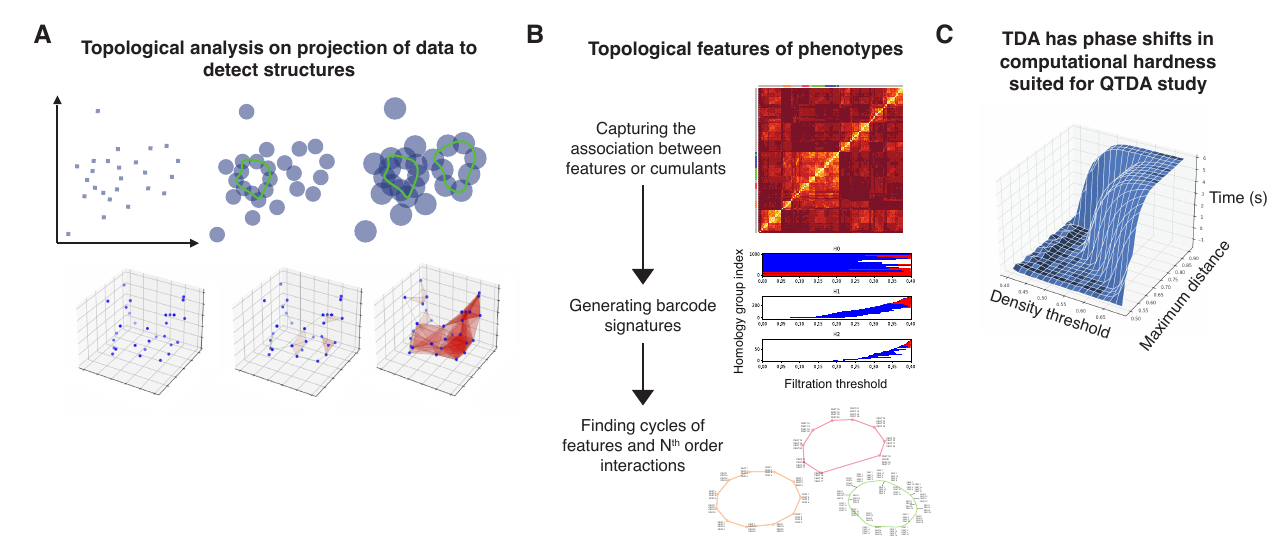}
    \caption{\textbf{Computational Topology for learning cellular behavior.}\\
    \textbf{A}: Examples of simplicial complexes detected in multidimensional data with each increase in the filtration distance (left to right) represented as increasing radii from the data points (top row) or new edges (red) drawn between points (bottom row). \textbf{B}: For a multinomal dataset, n-th order interactions and topological signatures can be discovered that define different phenotypes. The association between features or cumulants with some distance measure, e.g. pairwise Jaccard distances between features (heatmap). Applying TDA to the projection of data in that defined distance space generates topological features such as barcode signatures or cycles containing n-th order interactions that describe known and novel phenotypes in the data. \textbf{C}: Phase shifts in the computational time (z-axis) for classical TDA when testing combinations of density threshold (x-axis) and maximum distance (y-axis) when computing Vietoris-Rips complexes indicates there are specific problems spaces where quantum computing has opportunity to provide significant speed ups over classical TDA.\\
    }
    \label{fig:WS4overview}
\end{figure}

Topological data analysis and cumulants can capture these n-th order interactions that would represent learned complex associations leading to logical relationships explaining a given phenotype, like interactions between the community of cells/features/alterations that lead to therapeutic response (Fig.~\ref{fig:WS4overview}). 
As described earlier (see Section \ref{section:qml}), TDA discovers hidden structures in the shape of data to gain novel insights (Fig.~\ref{fig:WS4overview}A).
TDA is already being used in computer vision and atomistic models, and cumulants of order $d > 3$ are playing important roles in financial data analyses, economics, hyper-spectral image analyses, etc. Yet, they are classically constrained as to how far they can scale in terms of dimension ($d <5 $) and feature space ($< 100$).
Cumulants are a useful tool for integrating and analysing multi-omics data, whether single-cell or patient omics data, where they can provide a unique measurement of liability or risk of a given disease or trait, for each cell or individual (Fig.~\ref{fig:WS4overview}B). The association, i.e. distance, between features or cumulants can be calculated and  TDA applied to generate the topological features, e.g., barcode signatures or cycles, that characterize the clusters of features. The utility of this type of mining was demonstrated by constructing aggregated phenotypes representing distinct pathways to identify genetic variants relevant to that pathway via genome-wide association studies \cite{platt_characterizing_2016}, exploring metabolic syndrome with coronary artery disease, and metabolic interactions leading to COVID-19 severity, where it was found that renin angiotensin aldosterone system (RAAS) drugs mediated the risk of severe COVID-19 due to hypertension, and that lipids mediated several other metabolic syndrome risk factors for severity \cite{platt_epidemiological_2022,karisani_topology_2022}.

Overall, TDA and cumulant calculation are powerful tools for analyzing healthcare data, allowing researchers and healthcare professionals to gain a deeper understanding of complex medical conditions and improve patient outcomes. However there are classical bottlenecks in cumulant computation owing to the combinatorial expansions of the multinomial feature combinations, and in TDA computations where calculating exact and approximate Betti numbers is $\#P$ and $NP$-hard\cite{schmidhuber2022complexitytheoretic}, respectively. For instance, in most analyses of persistent Betti number calculations, there is a phase shift in the computational time needed to perform the calculation (Fig.~\ref{fig:WS4overview}C). This shift represents areas where quantum may offer particularly significant advantages over classical TDA. These classical constraints can be overcome using QTDA and quantum computing for cumulant computation. 

Quantum computing offers the potential for moving past the classical limitations for these techniques enabling more sophisticated applications of TDA and cumulant analysis, and other mathematical techniques to improve biomedical data analysis.  Hereto, quantum walks may also prove useful on the discovered knowledge graphs, as they have already been shown to be able to infer missing links in protein-protein interaction networks~\cite{Goldsmith2023}. By applying these techniques to the spatial single cell omics data, it would be possible to capture those cryptic interactions that would have been hidden even with complete biological measurements let alone incomplete ones encountered in practice to better inform spatial and perturbation models described above.

\section{Conclusion}
A cell-centric therapeutic design strategy can provide clinicians and patients with much needed additional treatment options, bringing us significantly closer to precision medicine. By developing a deeper understanding of and modeling how cancer cells behave individually and in aggregate, treatment plans can be developed to manipulate the cancer and its tumor microenvironment into a more therapeutically responsive state or shift the tumor into an indolent phase transforming the disease into a more manageable, chronic condition. Quantum computing is a powerful enabling technology to help push this approach to design therapeutics forward, and this case study may serve as an exemplar of how quantum computing can meaningfully contribute to HCLS.

\section{Acknowledgments}
Portions of this material is based upon work supported by NSF Center for Cellular Construction DBI-1548297 grant to S.C.

\bibliographystyle{unsrt}
\bibliography{main}

\end{document}